\documentstyle[12pt,axodraw]{article}
\input{psfig.tex}

\begin{document}
\begin{titlepage}

\hskip 12cm 
\rightline{\vbox{\hbox{DFPD 98/TH 51}\hbox{CS-TH 6/98}
\hbox{Dec. 1998}}}
\vskip 0.6cm
\centerline{\bf PHOTOPRODUCTION OF HEAVY VECTOR MESONS AT HERA --}
\centerline{\bf A TESTFIELD FOR DIFFRACTION$^{~\diamond}$}
\vskip 1.0cm
\centerline{  R. Fiore$^{a\dagger}$, L. L. Jenkovszky$^{b\ddagger}$, F.
Paccanoni$^{c\ast}$}
\vskip .5cm
\centerline{$^{a}$ \sl  Dipartimento di Fisica, Universit\`a della Calabria,}
\centerline{\sl Istituto Nazionale di Fisica Nucleare, Gruppo collegato di
Cosenza}
\centerline{\sl Arcavacata di Rende, I-87030 Cosenza, Italy}
\vskip .5cm
\centerline{$^{b}$ \sl  Bogoliubov Institute for Theoretical Physics,}
\centerline{\sl Academy of Sciences of the Ukrain}
\centerline{\sl 252143 Kiev, Ukrain}
\vskip .5cm
\centerline{$^{c}$ \sl  Dipartimento di Fisica, Universit\`a di Padova,}
\centerline{\sl Istituto Nazionale di Fisica Nucleare, Sezione di Padova}
\centerline{\sl via F. Marzolo 8, I-35131 Padova, Italy}
\vskip 1cm
\begin{abstract}
Exclusive diffractive photoproduction of heavy vector mesons (V=$\phi,
J/\psi$ and $\Upsilon$) at HERA is studied in a model employing a dipole 
Pomeron exchange (P) with an inelastic $\gamma PV$ vertex. The model is fitted 
to the data on ${d\sigma/dt}$, $B$ and $\sigma_{el}$ for $Q^2=0$ and 
beyond the threshold region. The elastic cross sections for both $\phi$ and
$J/\psi$ photoproduction show a moderate increase within the HERA energy 
region. The flattening of the slope $B(s)$ (little or no shrinkage) for 
$J/\psi$ is not correlated with the slope of the Pomeron trajectory. 
Estimates for $\Upsilon$ photoproduction at HERA are given.
\end{abstract}
\vskip .5cm
\hrule
\vskip .3cm
\noindent

\noindent
$^{\diamond}${\it Work supported by the Ministero italiano
dell'Universit\`a e della Ricerca Scientifica e Tecnologica and by the INTAS}
\vfill
$\begin{array}{ll}
^{\dagger}\mbox{{\it email address:}} &
   \mbox{FIORE~@CS.INFN.IT}
\end{array}
$

$ \begin{array}{ll}
^{\ddagger}\mbox{{\it email address:}} &
 \mbox{JENK~@GLUK.APC.ORG}
\end{array}
$

$ \begin{array}{ll}
^{\ast}\mbox{{\it email address:}} &
   \mbox{PACCANONI~@PADOVA.INFN.IT}
\end{array}
$
\vfill
\end{titlepage}
\eject
\textheight 210mm
\topmargin 2mm
\baselineskip=24pt

{\bf 1. INTRODUCTION}

\vskip 1.5 cm

Diffractive photoproduction, as well electroproduction, of heavy vector mesons 
at HERA continues attracting attention of both theorist and experimentalists 
(for a recent review see e.g. Refs.~\cite{critt1}- 
\cite{abbien}; a comprehensive earlier review on this subject can be found in 
Ref.~\cite{bauer}) as a unique testfield for diffraction,  
an interface between "soft" and "hard" physics, with three independent 
kinematical variables, the c.m.s energy $W=\sqrt s$, the transferred 
momentum $\sqrt {-t}$ and the virtuality of the external particle(s) $Q^2=-q^2$ 
involved simultaneously. High masses $M_V$ of the external vector mesons 
are usually treated on the same footing as the photon virtuality, 
by introducing the variable $\tilde Q^2=Q^2+M_V^2$, although $M^2_V$ should 
not be identified with $Q^2$. In the present paper we consider  only 
photoproduction, i.e. $Q^2=0$.  

An important reason why heavy vector mesons are particularly suitable to 
study diffraction is that, by the OZI rule \cite {OZI},  
photoproduction of heavy vector mesons is mediated by the exchange of 
a Regge trajectory with vacuum quantum numbers and made of gluons (the 
Pomeron trajectory). In the case of the $\phi$ production, a small contribution 
from subleading, secondary Reggeons -- due to the $\omega-\phi$ mixing is also 
possible. 

Reactions and/or kinematical regions 
with the Pomeron dominance (Pomeron "filters") have been looked for long
ago -- e. g. in the elastic scattering with exotic direct channels (like in 
$K^+ p$ or $pp$ scattering) or in other reactions (mainly $\bar p p$) at 
very high energies. In the hadron scattering, however, genuine Pomeron 
"filters" 
cannot be completely realized since even in the case of exotic channels a 
small contribution from secondary trajectories is inevitably present due to the 
breakdown of the exchange degeneracy. The alternative way to filter -- by 
going to 
very high energies -- is trapped by another (even less known) object -- the 
Odderon that obscures the picture and makes the discrimination ambiguous. In 
the case of photoproduction, only positive the C-parity exchange is allowed 
(the Odderon exchange is forbidden).                      

The application of the Regge pole theory to photoproduction usually implies 
also the validity of vector the meson dominance (VMD), by which the
photon, before interacting with the proton by means of a Reggeon exchange, 
first fluctuates, becoming a vector meson (Fig 1). The applicability of 
VMD and its generalizations to heavy meson states have been recently 
discussed in a number of papers \cite{schild}, \cite{HK}.

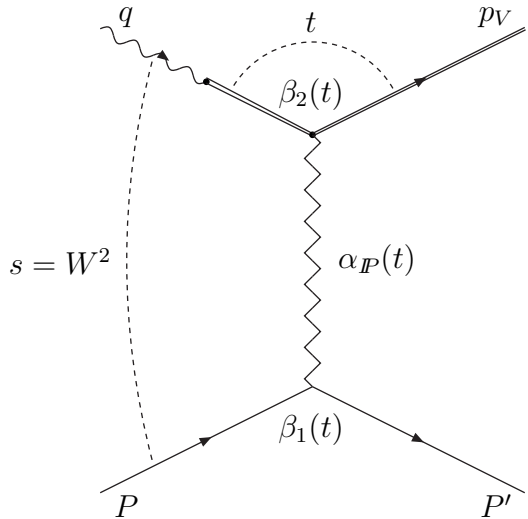
\begin{figure}
\begin{center}
\begin{picture}(300,200)(-10,-44)
\Text(50,145)[]{$q$}
\Text(190,145)[]{$p_V$}
\Photon(40,140)(80,120){2}{4}
\ArrowLine(63.2,130)(65.2,128.5)
\GCirc(80,120){1}{0}
\Line(80,119.5)(120,99.5)
\Line(80,121.5)(120,100.5)
\DashCArc(120,100)(35,30,148){2}
\Text(120,143)[]{$t$}
\Text(120,116)[]{$\beta_2(t)$}
\GCirc(120,100){1}{0}
\Line(120,99.5)(200.3,139.6)
\Line(120,100.5)(200,140.5)
\ArrowLine(160,120)(162,121.1)
\ZigZag(120,100)(120,5){3}{8}
\Text(145,53)[]{$\alpha_{I\!\!P}(t)$}
\ArrowLine(40,-35)(120,5)
\ArrowLine(120,5)(200,-35)
\Text(120,-11)[]{$\beta_1(t)$}
\Text(50,-40)[]{$P$}
\Text(190,-40)[]{$P'$}
\Text(25,52.5)[]{$s = W^2$}
\DashCArc(350,52.5)(300,165.5,194.5){2}
\end{picture}
\end{center}
\caption[]{\small Elastic photoproduction according to Vector Meson 
Dominance}
\label{fig1}
\end{figure}

An alternative to this typical Regge pole model of photoproduction, is  
perturbative QCD (pQCD). While (pQCD) calculations are efficient 
(see e.g. Refs.~\cite{ryskin}- 
\cite{RRML}) in the evaluation of the upper vertex of Fig. 2, or of the 
proton structure function probed by a "hard" Pomeron and related to the 
imaginary part of the photoproduction forward scattering , they are less 
appropriate for studying the typically non-perturbative features of 
diffraction, such as the $t$-dependence (shape of the assumed cone), the energy 
dependence of its slope, of the cross sections etc. 

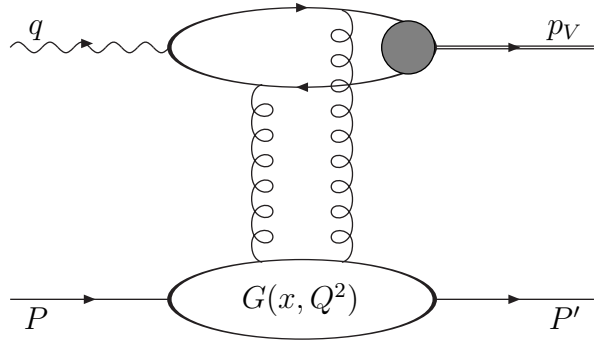
\begin{figure}
\begin{center}
\begin{picture}(200,145)(85,-10)
\Text(70,107)[]{$q$}
\Text(270,107)[]{$p_V$}
\Photon(60,100)(120,100){2}{5}
\ArrowLine(88,101.5)(90,101.5)
\ArrowLine(169,115)(171,115)
\Oval(170,100)(15,50)(0)
\ArrowLine(171,85)(169,85)
\GCirc(210,100){10}{0.5}
\Line(220,99.5)(280,99.5)
\Line(220,100.5)(280,100.5)
\ArrowLine(249,100)(251,100)
\Gluon(185,114)(185,19){4}{9}
\Gluon(155,19)(155,86){4}{6}
\Oval(170,5)(15,50)(0)
\Text(170,5)[]{$G(x,Q^2)$}
\ArrowLine(60,5)(120,5)
\ArrowLine(220,5)(280,5)
\Text(70,-2)[]{$P$}
\Text(270,-2)[]{$P'$}
\end{picture}
\end{center}
\caption[]{\small Elastic photoproduction according to perturbative QCD}
\label{fig2}
\end{figure}

Recent studies \cite{NNP} involve -- apart from the abovementioned 
measurable quantities -- also more subtle details, 
such as photoproduction of radially 
excited states, the helicity dependence  etc.  In a different paper \cite{LM-G} 
a detailed analysis of the $t-$ dependence of the cone, including a possible
dip-bump structure, seen in hadronic reactions was studied.   

Most of the existing models rely on the so-called two-component picture,
a compilation of the "soft" mechanism (see Fig. 1), 
essentially based on VMD and 
the Donnachie-Landshoff (DL) model of the Pomeron \cite{DL0} and a "hard" 
one based on pQCD calculation of the exchange 
of a pair of gluons coupled to the quark-antiquark pair, as illustrated in 
Fig. 2, or the dipole picture \cite{NNP} (not to be confused 
with the dipole Pomeron!), where the 
nonperturbative effects from the propagator are plugged into the vertices. 
Apart from the peculiarities of the different models, a common feature and main 
result of all these approaches is a power increase in energy of the cross 
sections $s^\epsilon,$ fed in from the DL model, the fitted value 
$\epsilon$ being considered indicative of the "hardness" of diffraction. 
Another argument in favour of the "hardness" of diffraction at HERA, widely 
discussed now in literature \cite{levy} is the apparent flatness 
(small, or vanishing slope) of the Pomeron trajectory. Anticipating our 
forthcoming discussion, here we only notice that the Pomeron intercept is 
universal, independent of the virtuality or mass of the external particles,
so the abovementioned effect may have a different origin.

The aim of the present paper is an analysis of the basic assumptions 
behind the existing models. To this end we use a factorized model with a 
Pomeron exchange combining Figs. 1 and 2, without specifying 
the details (VMD or pQCD) of the upper vertex. Instead, we consider 
a general form for the (inelastic) $\gamma P V$ vertex and a two-term Pomeron 
exchange (simple and double pole). By confronting the model with the data we 
study its physical consequences.     

\vskip 1.5 cm

{\bf 2. KINEMATICS AND THE HERA DATA}

\vskip 1.5 cm

Here we introduce the kinematics and make several general comments concerning 
the HERA data, both from ZEUS and H1 collaborations. 

We use the standard notation for the reaction energy (see Fig. 3). The 
square of the c.m.s. energy and the momentum transfer to the proton are, 
respectively,
\begin{displaymath}
W^2 = (q + P)^2~,~~~~~~t=(P - P')^2~, 
\end{displaymath}
being  
\begin{displaymath}
\mid t\mid_{min} \approx m_p^2{(M_V^2+Q^2)^2\over{W^4}}~.
\end{displaymath}
Here $M_V$ is the vector-meson mass, $m_p$ the proton mass and $Q^2=-q^2$ 
is the photon virtuality. In the following we will use the symbol $s$ to 
indicate $W^2$. 
At HERA one has $20~GeV < W < 240~GeV, \  \ 
-13~GeV^2 < t < -\mid t\mid_{min}$, with $\mid t\mid_{min}\approx 
10^{-4}~GeV$ negligibly small.

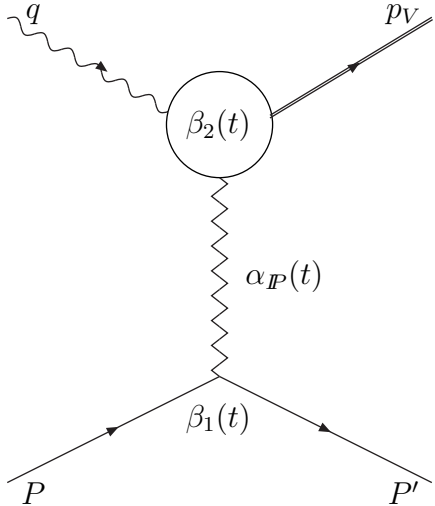
\begin{figure}
\begin{center}
\begin{picture}(200,200)(35,-40)
\Text(50,142)[]{$q$}
\Text(190,142)[]{$p_V$}
\Photon(40,140)(101,105){2}{6}
\ArrowLine(75,122)(77,120.5)
\GCirc(120,100){20}{1}
\Text(120,100)[]{$\beta_2(t)$}
\Line(139,102.5)(200.3,139.6)
\Line(139,103.5)(200,140.5)
\ArrowLine(170,121.5)(172,122.8)
\ZigZag(120,80)(120,5){3}{8}
\Text(145,43)[]{$\alpha_{I\!\!P}(t)$}
\ArrowLine(40,-35)(120,5)
\ArrowLine(120,5)(200,-35)
\Text(120,-11)[]{$\beta_1(t)$}
\Text(50,-39)[]{$P$}
\Text(190,-39)[]{$P'$}
\end{picture}
\end{center}
\caption[]{\small Elastic photoproduction with an inelastic
$\gamma P V$ vertex. The wiggle line, showing the Pomeron exchange, corresponds 
to a sum of two diagrams, i.e. simple and double pole exchanges.}
\label{fig3}
\end{figure}

     Since the differential cross section is the only directly 
observable quantity, $\sigma_{el}$, the slope $B$ and other quantities being 
derivatives, 
its determination and interpretation is of great importance; small errors 
in ${d\sigma/{dt}}$ may be amplified in $\sigma_{el}$ or 
in $B$. It should be admitted that the precision of the data is 
inferior to those in elastic hadron scattering. Therefore, in studying 
universal diffractive phenomena (such as the shape of the cone) or parameters 
(e.g. of the Pomeron trajectory) one should rely on the existing experience in 
hadronic (e.g. $p p$ or $\bar p p$) scattering at high energies. 
In particular, two unmistakable structures superimposed on the nearly 
exponential cone are known to exist (see e.g. Refs.~\cite{VJS}, \cite{DGJ}):

1. The "break", or changes of the local slope at $t\approx -0.1~GeV^2$, due to 
the nearby 2-pion threshold in the unphysical region ($t > 0)$, resulting in 
the sharpening of the cone (increase of $B(t)$ towards $t=0$). 
This tiny effect is not yet observable at the level of statistics 
typical of the HERA measurements. 

2. The dip-bump structure, clearly seen and thoroughly studied 
\cite{VJS} in hadronic reaction, is highly indicative of the diffractive 
phenomena. Its 
position, in general, is determined \cite{VJS} by the slope $B$ and the amount 
of absorptions. While the smaller -- with respect to the pp scattering -- 
slope in the heavy vector meson production evidently pushes the dip outwards,
the amount of absorptions is less known (it is expected to have a counter 
effect on the position of the dip). More data are needed to reveal the 
existence of a dip, which would be an important step towards a better 
understanding of diffraction.

The apparent flattening of the cone in $J/\psi$ photoproduction may seem
an indication of a nonlinear Pomeron trajectory. Here again the lesson from 
$p p$ and $\bar p p$ scatterings may be useful. 
It tells us \cite{VJS} that the slope 
of the Pomeron trajectory, apart from the small $\mid t \mid$ curvature due 
to the lightest two-pion threshold in the cross channel of the 
amplitude, remains almost constant until about $1~GeV^2$ -- the 
neighbourhood of the dip. The nonlinearity of the (Pomeron) trajectory can 
be of fundamental importance at large $\mid t \mid$, 
however the present HERA data are unlikely to tell us 
more about its details than the hadron scattering 
data do. Instead, the form of the nonlinear Pomeron trajectory gained 
from $pp$ and $\bar p p$ data may be used in identifying new effects at HERA. 

Given the abovementioned uncertainties, the formula 
\begin{displaymath}
\left. \sigma_{el}={1\over {B_{exp}}}{d\sigma\over{dt}}\right |_{t=0}~,
\end{displaymath}
where $B_{exp}$ is the experimental value of the slope, may be the right 
approximation. Formally it implies $B(t=0)$, although the slope 
can be determined only with respect to a finite interval (bin) in $t$. 
In view of the apparent flattening of the cone and the uncertainties in the 
determination of $d\sigma/dt$, the choice of the relevant bins and the 
resulting $B$ strongly influences the calculated $\sigma_{el}$. A reasonable 
way \cite{NNP} to account for the above-mentioned effect of the "sharpening" 
of $d\sigma/dt$ towards $t=0$, is by augmenting the measured $B$ by 1 
or 2 units of $GeV^2$. Since the determination of $B$ is crucial for the
dynamics of $J/\psi$ production (see Fig. 4 and below), this point needs 
further clarification. 
 
\vskip 1.5 cm

{\bf 3. DIFFRACTION AND REGGE POLE MODELS}

\vskip 1.5 cm

Factorization is a basic ingredient of any Regge pole model (see Ref.~
\cite{collins}). Accordingly, the scattering amplitude corresponding to a 
simple Regge pole exchange, up a signature factor $\xi(t),$ is a product of 
two vertices $\beta_1(t), \ \beta_2(t)$ and a "propagator" 
$(s/s_0)^{\alpha(t)}$ (see Fig. 1). If the amplitude is a sum of 
several exchanges (the Pomeron itself may be more than just a simple 
pole!), then each term conserves its factorization properties separately.     
      
The lower vertex in Fig. 1 is well known (from the $p p$ and $\bar p p$ 
scattering) to be ${\rm e}^{bt}$ (the application of more involved forms is
not relevant here), an estimate for $b$ being \cite{collins} 
$b=2.25~GeV^{-2}$. By factorization, the properties and values 
of the parameters in the Pomeron trajectory are universal and
reaction-independent. Below we use the "canonical" value of 
$\alpha'=0.25~GeV^{-2}$ for the Pomeron slope. This input is 
sufficient to calculate the slope of the exponential cone    
\begin{displaymath}
B(s)={d\over{dt}}\ln{d\sigma\over{dt}}\left. \right|_{t=0}~.
\end{displaymath}
As a result, for the extreme case of a point-like coupling in the upper 
vertex, $b_2=0$: 
\begin{equation}
B(s)=(4.5+0.5ln (s/s_0))~GeV^{-2}~, 
\label{z1}
\end{equation}
and one gets $B=8.75~GeV^{-2}$ at $W=70~GeV$ (with $s_0=1~GeV^2$) -- much 
too much compared to the data. This value can be lowered by: lowering 
$b$ \cite{DL1} and \cite{HKK} (difficult, since the above value is already a 
conservative estimate!), lowering 
$\alpha'$ (incompatible with factorization) and/or increasing 
$s_0$. The last option is acceptable, moreover demanded by the data on the 
hadron scattering \cite{VJS},\cite{DGJ}, although in the relevant fits the 
increase of $s_0$ is accompanied by a corresponding increase of $b$, the net 
effect for $B$, eq (1), remains nearly zero. Anyway, even the lower limit of 
$B$ according to Eq.~(1), $B=4.5~GeV^{-2}$, saturates the experimental value 
for the $J/\psi$ photoproduction (see Fig. 4). 
In other words, in the simple Regge
pole model, there is no room left for the radius of a vector meson as heavy 
as $J/\psi$ or $\Upsilon$.

The next important issue is whether the Pomeron contribution can be 
adequately represented by a single pole exchange -- as it is in the case 
in the DL model. If so, the total cross section is simply 
$\sigma_{t}\sim s^\epsilon$ and the elastic cross section is also a single 
power, $\sigma_{el}\sim s^{\epsilon}$.

\begin{figure}
\centerline{\psfig{figure=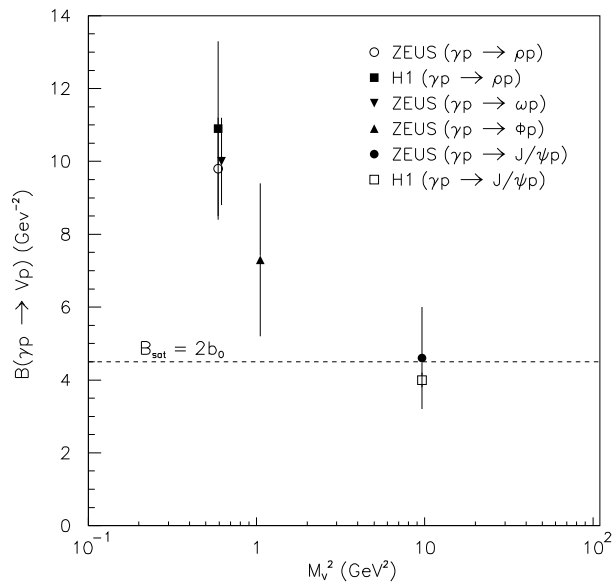,width=7.cm}}
\caption[]{\small The slope $B$ versus the square $M^2_V$ of the mass of 
Vector Bosons. Existing experimental data converge, saturating at the 
minimal value given by the lower $p P p$ vertex, $B=4.5~GeV^{-2}.$}
\label{fig4}
\end{figure}

The original DL model was always fitted to the data with two terms, the 
"Pomeron" and an effective contribution of subleading Reggeons. The latter
effectively contains also the low energy background (although this was never 
emphasized). The increasing part of the Pomeron should be added by a 
constant background - as it follows from the empirical fits to the total 
cross sections \cite{dino}, \cite{DGJ}, \cite{DGLM}, structure functions 
and their evolution \cite{JLP} or from non-perturbative QCD calculations 
\cite{JKP} .

To illustrate the aforesaid, let us write the Regge dipole scattering 
amplitude in a simple "geometrical" form (see \cite{VJS}):
\begin{displaymath}
A(s,t)\sim R^2{\rm e}^{R^2t}~,
\end{displaymath}
where $R^2\equiv R^2(s)=\alpha'(b+L-i\pi/2), \  L\equiv\ln{({s\over {s_0}})}$.
The total cross section in this model is
$$\sigma_{t}\sim b + L,$$ 
while the elastic cross section 
grows as
\begin{displaymath}
\sigma_{el}\sim{(b+L)^2+\pi^2/4\over{b+L}}
\end{displaymath}
and its asymptotics is delayed with respect to a single rising term (be it a 
power or logarithm(s)). By this simple example (anticipating a more 
realistic model to be presented in the next section), we intend to 
demonstrate the important role of a constant background to the rising 
term (whatever its form), and that the parametrizations of the 
HERA data by a single power $W^\epsilon$ may be oversimplified.     

\vskip 1.5 cm

{\bf 4. DIPOLE POMERON MODEL OF DIFFRACTION AT HERA}

\vskip 1.5 cm

We consider the reaction $\gamma p \rightarrow V p,$ where $V$ stands for 
$\phi,\ J/\psi$ or $\Upsilon$, in the framework of the Regge pole model 
with a dipole Pomeron (DP for brevity) exchange in the $t$ channel and 
inelastic $\gamma P V$ upper vertex shown in Fig. 3.

In the angular momentum plane, the partial wave amplitude corresponding to 
a Regge dipole is 
$$a(j, t)={\beta(j,t)\over{[j-\alpha(t)]^2}}=
{d\over{d\alpha(t)}}{\beta(j,t)\over{j-\alpha(t)}},$$
where the function $\beta(j)$ is $t$-independent and non-singular at 
$j=\alpha(t).$

The above derivative automatically produces a $\ln s$ term in the 
scattering amplitude, providing thus for rising cross sections with  
Pomeron intercept equal to one, securing the unitarity bounds.
Let us notice also that within Regge type models, this is the fastest 
rise allowed by unitarity, since asymptotically $\sigma_{t}\leq B$ and the 
maximally allowed shrinkage here is $B\sim\ln s.$

The DP model was successively applied to hadronic 
reactions in describing both the $s$- and $t$-dependence (for a review of the 
DP model see ~\cite{VJS}). Around $W\sim 100~GeV$ the rate of increase of 
the cross sections numerically is close to that in the DL model, i.e. 
$\sim W^{2\epsilon}, with \ \epsilon\approx 0.08$, 
but conceptually they are quite 
different. Furthermore, the interference between two terms, that can be 
interpreted as contributions from a simple and a double pole produces a 
diffractive pattern in $t$, confirmed experimentally in hadronic reactions 
(see Refs.~\cite{VJS}, \cite{DGJ}).   

In what follows we apply the above concept to diffractive 
photoproduction of heavy mesons.   
Neglecting the spin, we write the invariant scattering amplitude corresponding 
to the exchange of a DP as
$$A(s,t) = i(-is/s_0)^{\alpha(t)-1}\{G_1(t)+G_2(t)[\ln (s/s_0)-i\pi/2]\}~,
\eqno (2)$$
where 
$$G_1(t)=A_1{\rm e}^{bt}(1+h_1t) \eqno(3)$$
and 
$$G_2(t)=A_2{\rm e}^{bt}(1+h_2t)-\gamma \eqno(4)$$
are the residua of the simple and double pole, respectively. $G_1(t)$
factorizes (see, for instance, Fig. 3) into a 
standard $p P p$ vertex $\sim {\rm e}^{b t}$, 
with $b=2.25~GeV^{-2}$ determined from the $p p$ scattering, 
and the $\gamma P V$ vertex, that we parametrize by a simple polynomial 
with a free parameter $h_1$ to be fitted to the data. Were VMD applicable 
to the upper vertex, one should expect $h_1$ to be small and positive 
(expansion of an exponential with a small slope (radius)). If however, 
$h_1$ turns our to be negative, this will indicate departure from VMD
with an increase of the upper vertex in $\mid t\mid$.      

The residue of the double pole $G_2(t)$ may be cast from $G_1(t)$ by 
an integration (see \cite{VJS}). Here we relax this rather stringent 
constrain that relates the values of $A_2$ to $h_2$ and $A_1$ to $h_1$, 
respectively, keeping only the integration constant $\gamma$ as another free 
parameter. If that relation will be confirmed by the data, it will be 
indicative of the hadronic nature of diffraction in photoproduction.   

We use a simple linear trajectory for the 
Pomeron $\alpha(t)=1 + 0.25 t.$ This linear trajectory may be  
replaced by a nonlinear one in future, more sophisticated fits to the data.

\begin{figure}
\centerline{\psfig{figure=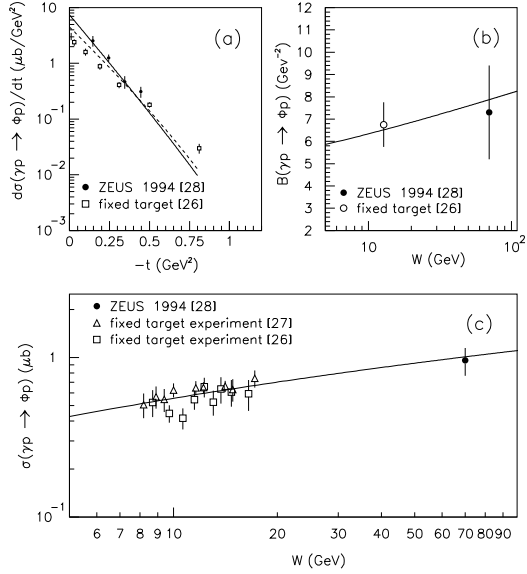,width=6.cm}}
\caption[]{\small $\phi$ photoproduction: a) the differential cross section:
the dotted line corresponds to the average energy $W= 13.3~GeV$ while the
full line to the average energy $70~GeV$; b) the slope parameter; c) the
elastic cross section. The data are taken from Refs.~\cite{busenitz},
\cite{alekhin} and \cite{HERA1}.}
\label{fig5}
\end{figure}

From Eq.~(1) we get for the elastic differential cross section 
$${d\sigma\over{dt}}=(s/s_0)^{2\alpha(t)-2}[(G_1(t)+G_2(t)\ln{(s/s_0)})^2+
{\pi^2\over 4} G_2^2(t)].\eqno(5)$$
whence $\sigma_{el}$ is calculated according to (see Sec. 2):
$$\left. {\sigma_el}={1\over B}{d\sigma\over{dt}}\right|_{t=0}.
\eqno(6)$$
The $s$ and $t$-dependence of the slope $B$ can be calculated from
$$\left. B(s,t)={d\over{dt}}(\ln{d\sigma\over{dt}})\right|_{t=o}=
2\alpha'\ln{(s/s_0)}+N/D,\eqno(7)$$
where 
$$N=2\bigg([A_1+(A_2-\gamma)\ln{(s/s_0)}][A_1(b+h_1)+
A_2(b+h_2)\ln{(s/s_0)}]+{\pi^2\over 4} (A_2-\gamma)A_2(b+h_2)\bigg)$$
and
$$D={d\sigma\over{dt}}\mid_{t=0}=~
[A_1+(A_2-\gamma)\ln{(s/s_0)}]^2+{\pi^2\over 4} (A_2-\gamma)^2.$$
In calculating $\sigma_{el}$ however we shall use the experimental value of 
the slope $B_{exp}$.

\vskip 1.5 cm

{\bf 5. FITS TO THE HERA DATA: DISCUSSION OF THE RESULTS} 

\vskip 1.5 cm

To study the Pomeron behaviour alone, unbiased by possible threshold effects, 
we impose a lower bound in energy, $W > 8~GeV$ in the case of the $\phi$ 
photoproduction and $W > 30~GeV$ in the case of the $J/\psi$ photoproduction.
As mentioned, here we consider only the case $Q^2=0$.

The parameters to be fitted are $A_1, A_2, h_1, h_2$ and $\gamma.$ 

Figs. 5 a) -- c) show the differential and the integrated elastic cross 
sections as well as the slope parameter of the $\phi$ production fitted to the 
fixed target \cite{busenitz}, \cite{alekhin} and the HERA collider \cite{HERA1} 
data. The values of the fitted parameters turn out to be $A_1 = 1.9126 \mu b, 
A_2 = 0.18203 \mu b,\  h_1=0.85842~GeV^{-2}. \ h_2=0$ and $s_0=(8~GeV)^2.$

Figs. 6 a -- c) show the same quantities for the 
$J/\psi$ photoproduction, with the 
fitted parameters: $A_1 = 0.27523 \mu b, A_2 = 0.091278 \mu b, \ \
h_1=-0.80606~GeV^{-2}, h_2=0$ and $s_0 = (30~GeV)^2$. Here only the HERA 
data \cite{HERA1} -- \cite{HERA3}  were used to fit the parameters.

\begin{figure}
\centerline{\psfig{figure=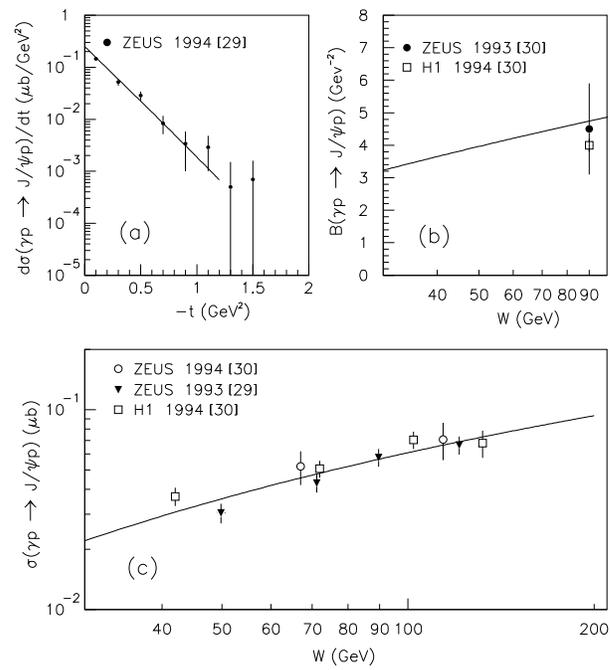,width=7.cm}}
\caption[]{\small $J/\psi$ photoproduction: a) the differential cross section; 
b) the slope parameter; c) the elastic cross section. The data are taken from 
Refs.~\cite{HERA1}, \cite{HERA2} and \cite{HERA3}.}
\label{fig6}
\end{figure}

The following comments are here in order

Throughout the fitting procedure $\gamma$ remains very small so, in order 
to reduce the number of the free parameters, we simple set $\gamma=0$. Since 
this parameter determines the amount of absorptions  (present in 
the model, see Ref.~\cite{VJS}) and the fate of a possible dip, this 
simplification is to be relaxed after a better understanding of the present
approach (from the previous experience in hadronic scattering 
\cite{VJS}, \cite{DGJ}, $\gamma$ is known to be small anyway).  

Photoproduction of $J/\psi$ requires the parameter $h_1$ to be negative. This 
is the effect of the "saturation" of the slope by the lower vertex only, 
visible in Fig. 4. To meet the data, the upper, inelastic vertex "subtracts" 
from the net slope. A negative value of $h$ does not favour VMD, 
indicating a more complicated, inelastic structure in the upper vertex of 
Fig. 3. 

We notice also that reasonable fits require large values of $s_0$, increasing 
with the mass of the produced vector meson. This parameter is correlated in 
some way with the external masses. Large values of $s_0\sim 100~GeV^2$ are 
typical also for the hadronic reactions \cite{VJS}, \cite{DGJ}, \cite{DL3},
\cite{WS}. 

Let us now discuss some general features in the behaviour of the 
observables, as they follow from our model. 

Even though we use a linear Pomeron trajectory, the cone is not exactly 
exponential due to the interference of the two terms of the Pomeron. 
An important immediate consequence is that the apparent non-shrinkage 
(little or no s-dependence in $B$) in the case of $J/\psi$ may 
result from the interference of the simple and double poles. Otherwise
stated, the form
$${d\sigma\over{dt}} = f(t) W^{[4\alpha(t)-4]},\eqno (8)$$
used in Ref.~\cite{levy} to fit the Pomeron trajectory and resulting in its 
apparent flatness, $\alpha'\approx 0,$ is not unique (e.g. $f$ may 
depend also on $s$). The flattening of  $B(s),$ visible in Fig. 6 c), 
can be achieved with a universal Pomeron intercept, $\alpha'=0.25~GeV^{-2}$ 
if e.g. (5) is used instead of (8). Moreover, the flattening of the slope may
be followed by an "antishrinkage" in $\Upsilon$ production (the negative, 
albeit small value of $\alpha'$ in \cite{levy} could be a message of this
trend).   

The energy dependence of the elastic cross sections,
shown in Figs. 5 a) and 6 a), 
is mild and fits the data perfectly well. The large mass of $J/\psi$ does 
not "harden" the dynamics, i.e. the rate of increase is similar to the case 
of its lighter counterpart $\phi$. Moreover, the present increase corresponds 
to a transitory regime, preceeding the asymptotic $\sim ln s$ rise, to set up
at still higher energies. 

To make predictions, we try to establish regularities between the values of the 
fitted parameters. Since the radius of the heavier $\Upsilon$ is smaller 
than that of $J/\psi$, which is already near, or even below the "saturation"
value (see Fig. 4), the effect 
of the "subtraction" (negative value of $h_1$) in the 
case of $\Upsilon$ is expected to be the same, or even weaker than in $J/\psi$.
So it may be reasonable to choose $h_{1,2}$ to be the same as in $J/\psi$.
The slope $B$ in the $\Upsilon$ photoproduction is also expected to be equal 
(or even slightly larger than) the saturation value $\approx 4.5~GeV^{-2}.$
The parameter $s_0$ tends to rise as $\ln M^2_V$, so by
extrapolation we choose $s_0 = (50~GeV)^2$ for $\Upsilon$ production. 

The relative normalization scale between the cross sections for various 
vector mesons is determined mainly by the parameter $A_1$ and it can be 
estimated according to the formula
$$A\sim (m_q)^{-4} M_V \Gamma_{V\rightarrow e^+ e^-} (e)^2,$$
where $M_V$ and $m_q$ are the masses of the relevant vector mesons and the
quark they contain, $\Gamma_{V\rightarrow e^+ e^-}$ is the decay width of 
the vector meson and $e$ is the electric charge of the relevant quark. It 
reproduces qualitatively the ratio between $\phi, \ \ J/\psi$ and $\Upsilon$
photoproduction. Since, however, the parameter $A_2$ also contributes to the
relative scales, we better rely on the recently measured $\Upsilon$ 
photoproduction cross section \cite{Upsilon} to fit these parameters.    

Setting $\gamma=0$, we adjust uniquely the normalization constant $A_1$ to 
the measured value \cite{Upsilon} of $\sigma_{el}$ in the $\Upsilon$
photoproduction and get $0.2 ~pb$. The parameter $A_2$, on the other hand 
shows more flexibility, by varying for fixed $A_1$ within the range 
$0.1\leq A_2 \leq 0.37 ~pb$, the central value being $A_2=0.2 ~pb$. 
The predicted elastic cross section is shown in Fig. 7.  

\begin{figure}
\centerline{\psfig{figure=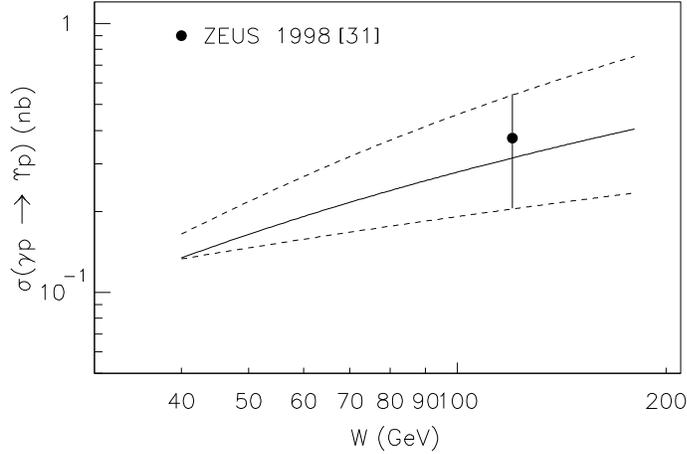,width=8.cm}}
\caption[]{\small A prediction for the elastic $\Upsilon$ photoproduction. 
The data point is taken from Ref.~\cite{Upsilon}.}
\label{fig7}
\end{figure}

Finally, we note that the present fits -- because of 
to the limited number of the data 
points relative to the number of the free parameters -- should be considered 
as preliminary, aimed at an exploration of a dynamical mechanism of 
diffractive photoproduction, alternative to the existing ones. Further 
comparison with the data may result in a different set of the fitted 
parameters, although the general features are expected to remain unchanged.

\vskip 1.5cm
\underline{Acknowledgements}:
L.Jenkovszky is grateful to the Dipartimento di Fisica dell'Universita' 
della Calabria, and to the Istituto Nazionale di Fisica Nucleare Sezione
di Padova and Gruppo Collegato di Cosenza, for the warm hospitality and 
financial support while part of this work was done.

\end{document}